\documentclass[11pt]{article} 

\usepackage[utf8]{inputenc} 

\usepackage{geometry} 
\usepackage{graphicx} 

\usepackage{booktabs} 
\usepackage{array} 
\usepackage{paralist} 
\usepackage{verbatim} 
\usepackage{subfig} 
\usepackage{color}
\usepackage{crayola}
\usepackage{hyperref}
\usepackage{listings}
\usepackage{amsmath}
\usepackage{amsfonts}
\usepackage{kbordermatrix}

\usepackage{fancyhdr} 
\usepackage{sectsty}
\allsectionsfont{\sffamily\mdseries\upshape} 

\usepackage[nottoc,notlof,notlot]{tocbibind} 
\usepackage[titles,subfigure]{tocloft} 

\newif\ifdraft \newif\ifblind
\draftfalse 
\blindfalse

\newcommand{\keyw}[1]{\textcolor{red}{\emph{#1}}}
\newcommand{\network}[1]{\mathcal{#1}}
\newcommand{\Hm}{\mathbf{H}}
\newcommand{\Net}{\network{N}}
\newcommand{\RR}{\Bbb{R}}
\newcommand{\NA}{\raisebox{-1pt}{$\square$}}
\newcommand{\wod}{\operatorname{wod}}
\newcommand{\wid}{\operatorname{wid}}
\newcommand{\od}{\operatorname{od}}
\newcommand{\id}{\operatorname{id}}
\newcommand{\bin}{\operatorname{bin}}
\newcommand{\LL}{\operatorname{L}}
\newcommand{\init}{\operatorname{init}}
\newcommand{\term}{\operatorname{term}}

\makeatletter
\newcommand{\firstof}[1]{\@car#1\@nil}
\newcommand{\secondof}[1]{\expandafter\@car\@cdr#1\@nil\@nil}
\newcommand{\restof}[1]{\expandafter\@cdr\@cdr#1\@nil\@nil}
\makeatother
\newcommand{\m}[1]{\firstof{#1}\!\!\secondof{#1}}
\newcommand{\s}[1]{\firstof{#1}\!\secondof{#1}}

\newcommand{\clock}{\count254=\time \divide\count254 by 60
 \count255=\count254 \multiply\count255 by -60
 \advance\count255 by \time
 \ifnum\count254<10 0\fi\number\count254\,:\,%
 \ifnum\count255<10 0\fi\number\count255}

\title{\bf Weighted degrees and truncated derived bibliographic networks}
\ifblind\author{Author\\ Institution\\ e-mail}\else
\author{Vladimir Batagelj\\IMFM Ljubljana and IAM UP Koper\\ 
e-mail: \texttt{vladimir.batagelj@fmf.uni-lj.si}\\ ORCID: 0000-0002-0240-9446}
\fi

\ifdraft\date{\today\  at \clock}\else\date{}\fi

\graphicspath{{./pics/}{./}}

\begin{document}
\ifblind\else
\hypersetup{pdfauthor={V. Batagelj}}
\hypersetup{pdftitle={Weighted degrees and truncated derived bibliographic networks}}
\fi

\maketitle

\begin{abstract}
Large bibliographic networks are sparse -- the average node degree is small. This is not necessarily true for their product -- in some cases, it can ``explode'' (it is not sparse, increases in time and space complexity). An approach in such cases is to reduce the complexity of the problem by limiting our attention to a selected subset of important nodes and computing with corresponding truncated networks.  The nodes can be selected by different criteria. An option is to consider the most important nodes in the derived network -- nodes with the largest weighted degree. It turns out that the weighted degrees in the derived network can be computed efficiently without computing the derived network itself. 
 \medskip

\end{abstract}

\section{Introduction}

From a selected bibliographic data we can construct a collection of corresponding bibliographic networks such as the authorship network $\Net_{\m{WA}}$, the keywords network  $\Net_{\m{WK}}$, the citation network $\Net_{Ci}$, etc. \cite{onbib}. For example, in the two-mode authorship network $\Net_{\m{WA}} = ((W, A), L_{\m{WA}})$ the set of nodes is split to the set of works $W$ and the set of authors $A$ and the set of links $L_{\m{WA}}$ consists of arcs (directed links) $(w, a) \in L_{\m{WA}}$ stating that the work $w \in W$ was co-authored by the author $a \in A$. The citation network $\Net_{Ci} = (W, L_{Ci})$ is a directed one-mode network on works with arcs $(w, z) \in L_{Ci}$ stating that the work $w \in W$ is citing the work $z \in W$.

In real-life networks, some nodes can be isolated. For example, in the authorship network, some works can have no author. An option would be to remove such nodes from the network. Because these nodes can have links in other collection's networks (for example, such a work can have links to the corresponding keywords in the keywords network) we decided to analyze the networks without removing isolated nodes. This leads to sometimes more complicated but also more general results.

Networks from a collection share some set of nodes. For example the authorship network $\Net_{\m{WA}}$ and the citation network $\Net_{Ci}$ share the set of works $W$. This allows us to compute, using network matrix multiplication, derived networks such as the co-authorship network $\mathbf{Co} = \mathbf{\m{WA}}^T \cdot \mathbf{\m{WA}}$ or the network of citations between authors $\mathbf{ACiA} = \mathbf{\m{WA}}^T \cdot \mathbf{Ci} \cdot \mathbf{\m{WA}}$ ($\mathbf{\m{WA}}^T$ is the transpose of matrix $\mathbf{\m{WA}}$) .

Large bibliographic networks are sparse -- the number of links is of the same order as the number of nodes (the average node degree is small). This is not necessarily true for their product -- in some cases, it can ``explode'' (it is not sparse, increases in time and space complexity)  \cite{onbib}. The problem with space (computer memory) can be dealt with using a sparse matrix representation and partitioning matrices into blocks. An approach to deal with the time complexity is to reduce the size of the problem by limiting our attention to a selected subset of important nodes and computing with corresponding truncated networks.  The nodes can be selected by different criteria. An option is to consider the most important nodes in the derived network -- nodes with the largest weighted degree. It turns out that the weighted degrees can be computed efficiently without computing the derived network itself. This idea is elaborated in this paper.


\section{Networks and product of networks}

In a \keyw{two-mode} (affiliation or bipartite) network $\Net = ((U,V),L,w)$ the set of nodes is split into two disjoint sets (\keyw{modes}) $U$ and $V$. The set of links can be described with the predicate $\LL(u,e,v) \equiv$ the link $e$ leads from the node $u$ to the node $v$. The node $u$ is called the \keyw{initial} node of the link $e$, $u \in \init(e)$, and the node $v$ is a \keyw{terminal} node of the link $e$, $v \in \term(e)$. The function $w : L \to \RR$ assigns to each link $e$ its \keyw{weight} $w(e)$. Many two-mode networks are binary -- the weight $w$ has a constant value 1 on all links.

In general, the weight can be measured on different measurement scales (counts, ratio, interval, ordinal, nominal, binary, TQ, etc.). Usually, a semiring structure is assumed on the set of its possible values. In this paper we will limit our discussion to the semiring $(\RR, +, \cdot, 0, 1)$ and its subsemirings \cite{semirings}.

In the case when $0$ is a possible value of the weight $w$, we extend the semiring of real numbers to the semiring $(\RR \cup \{ \NA \}, +, \cdot, \NA, 1)$ with a new zero $\NA$  with rules
$\NA + a = a$ and $\NA \cdot a = a \cdot \NA = \NA$. The value $\NA$ is interpreted as ``no link''.

The network \keyw{matrix} $\mathbf{M} = [ m[u,v] ]_{U\times V}$ of a two-mode network ${\cal N}$ is defined as
\[   m[u,v]  =  \left\{\begin{array}{ll} 
      \sum_{e \in L : \LL(u,e,v)} w(e) & 
          \quad \exists e \in L \exists u \in U \exists v \in V: \LL(u,e,v)\\
    \NA  & \quad \textrm{otherwise} 
     \end{array}\right.
\]
The \keyw{skeleton} or \keyw{binarized} version $ \bin(\mathbf{M}) = [  \delta[u,v]]_{U\times V}$ of the network matrix $\mathbf{M}$ is defined by
\[   \delta[u,v]  =  \left\{\begin{array}{ll} 
      1  & \quad m[u,v] \ne \NA \\
      \NA  & \quad \textrm{otherwise} 
     \end{array}\right.
\]
For a network $\Net $ with matrix $\mathbf{M}$,  we define its \keyw{total} $T(\Net) = T(\mathbf{M})$ as the sum of all its entries, 
\[T(\Net) = \sum_{e \in L} w(e) = \sum_{u \in U} \sum_{v \in V} m[u,v] =T(\mathbf{M}) \] 

For a matrix $\mathbf{M}_{U\times V}$, the matrix $\mathbf{M}^T = [m^T[v,u]]_{V\times U}$ is called its \keyw{transpose} and is determined by $m^T[v,u] = m[u,v]$. It holds $(\mathbf{M}^T)^T = \mathbf{M}$.
A square matrix $\mathbf{M}_{V\times V}$ is \keyw{symmetric} iff $\mathbf{M}^T = \mathbf{M}$. 

Some additional notions will be used in the following.
\begin{itemize}
\item the set of \keyw{out-neighbors} of the node $u \in U$:\\
$oN(u) = \{ v \in V : \exists e \in L : \LL(u,e,v) \}$
\item the set of \keyw{in-neighbors} of the node $v \in V$:\\
$iN(v) = \{ u \in U : \exists e \in L : \LL(u,e,v) \}$
\item the \keyw{out-degree} of the node $u \in U$: $ \od_{M}(u) = |oN(u)| $ 
\item the \keyw{in-degree} of the node $v \in V$: $ \id_{M}(v) = |iN(v)|  $
\item the \keyw{weighted out-degree}  of the node $u \in U$: $ \wod_{M}(u) = \sum_{v \in V} m[u,v] = \sum_{e : u \in \init(e)} w(e)$
\item the \keyw{weighted in-degree} of the node $v \in V$: $ \wid_{M}(v) = \sum_{u \in U} m[u,v] = \sum_{e : v \in \term(e)} w(e)$
\end{itemize}
In vector form, it holds
\[ \mathbf{wid}_{M^T} = \mathbf{wod}_{M} \qquad \mbox{and} \qquad \mathbf{wod}_{M^T} = \mathbf{wid}_{M} \]
\[ \mathbf{wid}_{\bin(M)} = \mathbf{id}_{M} \qquad \mbox{and} \qquad \mathbf{wod}_{\bin(M)} = \mathbf{od}_{M} \]

\begin{figure}
\includegraphics[width=45mm]{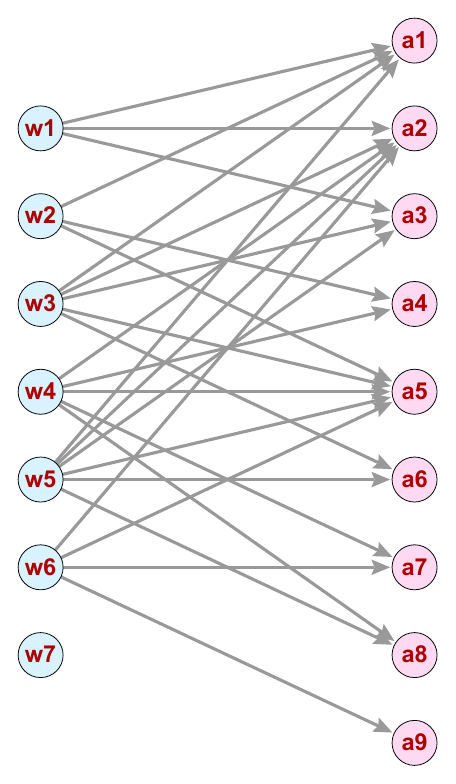}
\qquad \raisebox{35mm}{$ \mathbf{\m{WA}} = \kbordermatrix{
 & a1 & a2 & a3 & a4 & a5 & a6 & a7 & a8 & a9\\
 w1  & 1 & 1 & 1 & 0 & 0 & 0 & 0 & 0 & 0 \\
 w2  & 1 & 0 & 0 & 1 & 1 & 0 & 0 & 0 & 0 \\
 w3  & 1 & 1 & 1 & 0 & 1 & 1 & 0 & 0 & 0 \\
 w4  & 0 & 1 & 0 & 1 & 1 & 0 & 1 & 1 & 0 \\
 w5  & 1 & 1 & 1 & 0 & 1 & 1 & 0 & 1 & 0 \\
 w6  & 0 & 1 & 0 & 0 & 1 & 0 & 1 & 0 & 1 \\
 w7  & 0 & 0 & 0 & 0 & 0 & 0 & 0 & 0 & 0 \\
}$}

\includegraphics[width=45mm]{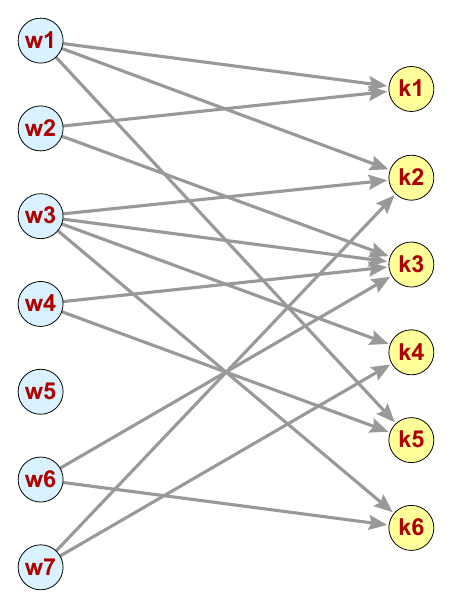}
\qquad \raisebox{27mm}{$ \mathbf{\s{WK}} = \kbordermatrix{
 & k1 & k2 & k3 & k4 & k5 & k6\\
 w1  & 1 & 1 & 0 & 0 & 1 & 0 \\
 w2  & 1 & 0 & 1 & 0 & 0 & 0 \\
 w3  & 0 & 1 & 1 & 1 & 0 & 1 \\
 w4  & 0 & 0 & 1 & 0 & 1 & 0 \\
 w5  & 0 & 0 & 0 & 0 & 0 & 0 \\
 w6  & 0 & 0 & 1 & 0 & 0 & 1 \\
 w7  & 0 & 1 & 0 & 1 & 0 & 0 \\
} $}

\includegraphics[width=40mm]{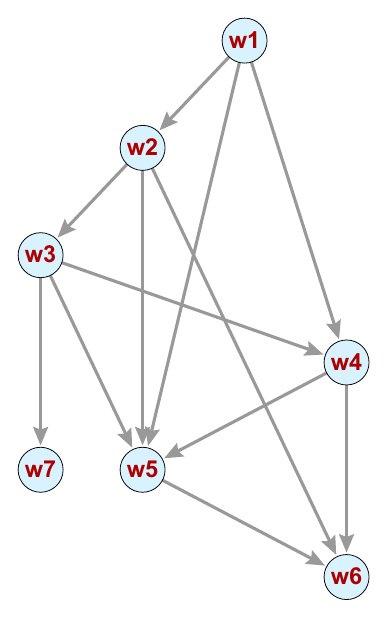}
\quad \ \ \qquad \raisebox{25mm}{$ \mathbf{Ci} = \kbordermatrix{
 & w1 & w2 & w3  & w4 & w5 & w6 & w7\\
 w1  & 0 & 1 & 0 & 1 & 1 & 0 & 0 \\
 w2  & 0 & 0 & 1 & 0 & 1 & 1 & 0 \\
 w3  & 0 & 0 & 0 & 1 & 1 & 0 & 1 \\
 w4  & 0 & 0 & 0 & 0 & 1 & 1 & 0 \\
 w5  & 0 & 0 & 0 & 0 & 0 & 1 & 0 \\
 w6  & 0 & 0 & 0 & 0 & 0 & 0 & 0 \\
 w7  & 0 & 0 & 0 & 0 & 0 & 0 & 0 \\
}$}

\caption{A toy network collection.\label{toy}}
\end{figure}

The \keyw{product} $\mathbf{C} = \mathbf{A} \cdot \mathbf{B} = [c[i,j]]_{I\times J} $ of two compatible  matrices $\mathbf{A}_{I\times K}$ and $\mathbf{B}_{K\times J}$ is defined in the standard way
\[  c[i,j] = \sum_{k \in K} a[i,k] \cdot b[k,j] \] 
It holds $(\mathbf{A} \cdot \mathbf{B})^T = \mathbf{B}^T \cdot \mathbf{A}^T $.

The product of two compatible networks ${\cal N}_A = ((I,K), L_A, a)$ and ${\cal N}_B = ((K,J), L_B, b)$ is the network ${\cal N}_C = ((I,J), L_C, c)$ where  $L_C = \{ (i,j) : c[i,j] \ne \NA \}$ and the weight $c$ is determined by the matrix ${\mathbf C}$, $c(i,j) = c[i,j]$. 

In binary networks $\network{N}_A$ and $\network{N}_B$, the value of $c[i,j]$ counts the number of ways we can go from the node $i \in I$ to the node $j \in J$ passing through $K$, $ c[i,j] = | oN_A(i) \cap iN_B(j)|$ where $oN_A(i)$ is the set of out-neighbors of the node $i\in I$ in the network ${\cal N}_A$ and  $iN_B(j)$ is the set of in-neighbors of the node $j\in J$ in the network ${\cal N}_B$.

For a matrix $\mathbf{M}_{U\times V}$, we define its \keyw{row projection} 
\[ r(\mathbf{M}) = \mathbf{M} \cdot \mathbf{M}^T \]
and its \keyw{column projection} 
\[ c(\mathbf{M}) = \mathbf{M}^T \cdot \mathbf{M} \]
We have $r(\mathbf{M}^T) = c(\mathbf{M})$. Both projections are symmetric
\[ r(\mathbf{M})^T = (\mathbf{M} \cdot \mathbf{M}^T)^T = \mathbf{M} \cdot \mathbf{M}^T = r(\mathbf{M}) \]
\[ c(\mathbf{M})^T = (\mathbf{M}^T \cdot \mathbf{M})^T = \mathbf{M}^T \cdot \mathbf{M} = c(\mathbf{M}) \]

For vectors  $\mathbf{x} = [x_1, x_2, \ldots, x_n]$ and $\mathbf{y} = [y_1, y_2, \ldots, y_m]$ their \keyw{outer product} $\mathbf{x} \circ \mathbf{y}$ is defined as a matrix
$$\mathbf{x} \circ \mathbf{y} = [x_i \cdot y_j]_{n\times m}$$
then we can express the product $\mathbf{C}$ of two compatible  matrices $\mathbf{A}$ and $\mathbf{B}$ as the \keyw{outer product decomposition} \cite{fraction}
$$  \mathbf{C} = \mathbf{A} \cdot \mathbf{B} = \sum_k \Hm_k  \quad \mbox{where} \quad \Hm_k  =  \mathbf{A}[\cdot,k] \circ \mathbf{B}[k,\cdot],  $$
where $\mathbf{A}[\cdot,k]$ is the $k$-th column of matrix $\mathbf{A}$,  and $\mathbf{B}[k,\cdot]$ is the $k$-th row of matrix $\mathbf{B}$.

Based on outer product decomposition, we have
\[  T(\mathbf{C}) = T(\sum_k   \Hm_k ) = \sum_k  T(\Hm_k )  \mbox{ \  and \  } 
T(\Hm_k ) = \mbox{wid}_A(k) \cdot \mbox{wod}_B(k) \]
and therefore
$ T(\mathbf{C}) =  \mathbf{wid}_A \cdot \mathbf{wod}_B $ ($\cdot$ denotes the inner/scalar product of vectors $\mathbf{x} \cdot \mathbf{y} = \sum_k x_k \cdot y_k$).

\subsection{Example}

To illustrate the introduced notions and check some of the derived relations we wrote a collection of R functions \texttt{\textbf{bibmat.R}} that support introduced operations on bibliographic networks represented by matrices  \cite{bibR}. For analyzing large networks we can use their implementation in Pajek as Pajek's commands or macros.

We also prepared a toy-example collection that contains an authorship network $\mathbf{\m{WA}}$, a  keywords network $\mathbf{\s{WK}}$, and a citation network $\mathbf{Ci}$ (see Figure~\ref{toy}).
Because all three networks are binary the corresponding degrees and weighted degrees vectors are equal
\[ \mathbf{wod}_{\m{WA}} = \mathbf{od}_{\m{WA}} = \kbordermatrix{
 & w1 & w2 & w3 & w4 & w5 & w6 & w7\\
 & 3 & 3 & 5 & 5 & 6 & 4 & 0\\
} \]
\[ \mathbf{wod}_{\s{WK}} = \mathbf{od}_{\s{WK}} = \kbordermatrix{
 & w1 & w2 & w3 & w4 & w5 & w6 & w7\\
 & 3 & 2 & 4 & 2 & 0 & 2 & 2\\
} \]
\[ \mathbf{wid}_{\m{WA}} = \mathbf{id}_{\m{WA}} = \kbordermatrix{
 & a1 & a2 & a3 & a4 & a5 & a6 & a7 & a8 & a9\\
 & 4 & 5 & 3 & 2 & 5 & 2 & 2 & 2 & 1\\
} \]
\[ \mathbf{wid}_{\s{WK}} = \mathbf{id}_{\s{WK}} = \kbordermatrix{
 & k1 & k2 & k3 & k4 & k5 & k6\\
 & 2 & 3 & 4 & 2 & 2 & 2\\
} \]

\section{Truncated derived networks}

\begin{figure}
\[\includegraphics[width=\textwidth]{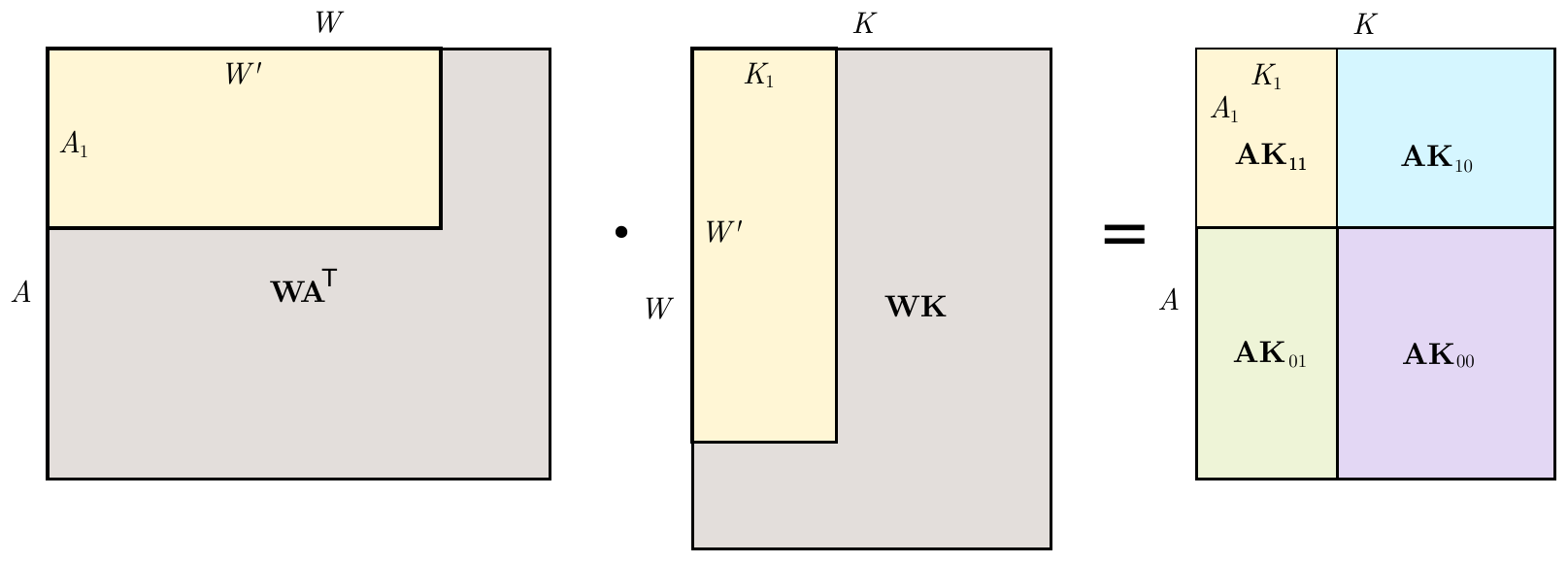}\]
\caption{Truncated derived network.\label{trunc}}
\end{figure}

Let's split the set of authors $A$ into two sets $A_1$ (selected authors) and $A_0$ (remaining authors), $A_1 \cup A_0 = A$ and $A_1 \cap A_0 = \emptyset$. We do the same with the set of keywords $K$ -- $K_1$ (selected keywords) and $K_0$ (remaining keywords), $K_1 \cup K_0 = K$ and $K_1 \cap K_0 = \emptyset$. We call a \keyw{truncated derived network} the network 
\[  \mathbf{\s{AK}}_{11} = \mathbf{\s{AK}}[A_1, K_1] = \mathbf{\m{WA}}^T_{W' \times A_1 } \cdot \mathbf{\s{WK}}_{W' \times K_1 } \]
where $W' = \{ w \in W : (\od_{\m{WA}_{W \times A_1}}(w) > 0) \land (\od_{\m{WK}_{W \times K_1}}(w) > 0) \}$.

For a selected author $a \in A_1$, we denote with $\mbox{oInt}(a) = \wod_{AK_{11}}(a)$ her/his \keyw{internal} out-contribution, and with $\mbox{oExt}(a) = \wod_{\s{AK}}(a) - \mbox{oInt}(a)$ her/his \keyw{external} out-contribution. And similarly, for a selected keyword $k \in K_1$,we denote with $\mbox{iInt}(k) = \wid_{AK_{11}}(k)$ its \keyw{internal} in-contribution, and with $\mbox{iExt}(k) = \wid_{\s{AK}}(k) - \mbox{iInt}(k)$ its \keyw{external} in-contribution. In the next Section, we shall show that weighted degrees of $\mathbf{\s{AK}}$ can be determined without computing the network itself.

We reorder the nodes of the network $\mathbf{\s{AK}}$ according to the $A_1$, $A_0$ and $K_1$, $K_0$ splits (see Figure~\ref{trunc}). The derived network matrix $\mathbf{\s{AK}}$ is split into four submatrices $\mathbf{\s{AK}}_{ij}, i,j \in \{0,1\}$. We denote their totals $T_{ij} = T(\mathbf{\s{AK}}_{ij})$.  $T_{11}$ is the contribution of cooperation among selected nodes, $T_{10} + T_{01}$ is the contribution of cooperation of selected nodes with remaining nodes, and $T_{00}$ is the contribution of cooperation among remaining nodes. We can compute all four totals
\[ T_{11} = T(\mathbf{\s{AK}}_{11}) = \sum_{a \in A_1} \mbox{oInt}(a) = \sum_{k \in K_1} \mbox{iInt}(k) \]
\[ T_{10} = \sum_{a \in A_1} \mbox{oExt}(a) \quad \mbox{and} \quad T_{01} = \sum_{k \in K_1} \mbox{iExt}(k) \]
and finally 
\[ T_{00} =  T(\mathbf{\s{AK}}) - T_{11} -  T_{10} -  T_{01}. \]
Note that we used only information from $\mathbf{\m{WA}}$, $\mathbf{\s{WK}}$ and $\mathbf{\s{AK}}_{11}$.

\section{Weighted degrees}

\subsection{Weighted degrees in derived networks}
Often in a derived network, the importance of its nodes is measured by their weighted degree. It turns out that we don't need to compute the derived network to get them. They can be determined faster. Let's consider a general case
\[  \mathbf{\s{AK}} = \mathbf{\m{WA}}^T \cdot \mathbf{\s{WK}} \]
where $ \mathbf{\m{WA}}$ and $\mathbf{\s{WK}}$ are compatible two-mode networks. The interpretation of the network $\mathbf{\s{AK}}$ depends on the nature of networks  $ \mathbf{\m{WA}}$ and $\mathbf{\s{WK}}$.

In the case when  $ \mathbf{\m{WA}}$ and $\mathbf{\s{WK}}$ are the authorship and the keywords matrices, the entry $ak[a,k]$ counts the number of different triples $(a,w,k)$ such that the author  $a$ wrote the work $w$ that is described by the keyword $k$ -- the number of times the author $a$ is dealing with the topic $k$ in his/her works.

For the weighted out/in-degrees of $\mathbf{\s{AK}}$ we get
\[ \wod_{AK}(a) = \sum_{k \in K} ak[a,k] = \sum_{k \in K} \sum_{w \in W} wa^T[a,w] \cdot wk[w,k] = \]
\[ =  \sum_{w \in W} wa[w,a] \cdot \sum_{k \in K} wk[w,k] =  \sum_{w \in W} wa[w,a] \cdot \wod_{\s{WK}}(w) \] 
or in a vector form
\[ \mathbf{wod}_{AK} = \mathbf{\m{WA}}^T \cdot \mathbf{wod}_{\s{WK}} \qquad \qquad (\mbox{Oeq}) \]
and
\[ \wid_{\s{AK}}(k) = \sum_{a \in A} ak[a,k] = \sum_{a \in A} \sum_{w \in W} wa^T[a,w] \cdot wk[w,k] = \]
\[ =  \sum_{w \in W} wk[w,k] \cdot \sum_{a \in A} wa[w,a] =  \sum_{w \in W} wk[w,k] \cdot \wod_{\m{WA}}(w) \] 
or in a vector form
\[ \mathbf{wid}_{AK} = \mathbf{\s{WK}}^T \cdot \mathbf{wod}_{\m{WA}} \qquad \qquad (\mbox{Ieq}) \]
and finally for the network total
\[ T(\mathbf{\s{AK}}) =  \sum_{a \in A} \sum_{k \in K} ak[a,k] = \sum_{w \in W}
 \sum_{a \in A} wa[w,a] \cdot \sum_{k \in K} wk[w,k] =  \]
\[ = \sum_{w \in W}
 \wod_{\m{WA}}(w) \cdot \wod_{\s{WK}}(w) =  \mathbf{wod}_{\m{WA}} \cdot \mathbf{wod}_{\s{WK}}\]
 From equalities (Oeq) and (Ieq) we see that indeed both weighted degrees of  $\mathbf{\s{AK}}$  can be computed faster.

\subsubsection{Example}
For our toy example, using equalities (Oeq) $\mathbf{wod}_{\s{AK}} = \mathbf{\m{WA}}^T \cdot \mathbf{wod}_{\s{WK}} $ and  (Ieq) $\mathbf{wid}_{\s{AK}} = \mathbf{\s{WK}}^T \cdot \mathbf{wod}_{\m{WA}} $, we compute vectors 
 \[ \mathbf{wod}_{\s{AK}} =
 \kbordermatrix{
 & a1 & a2 & a3 & a4 & a5 & a6 & a7 & a8 & a9\\
   & 9 & 11 & 7 & 4 & 10 & 4 & 4 & 2 & 2 \\
 } \]
 \[ \mathbf{wid}_{\s{AK}} =
 \kbordermatrix{
 & k1 & k2 & k3 & k4 & k5 & k6\\
   & 6 & 8 & 17 & 5 & 8 & 9 \\
} \]
We see that works $w1, w3, w4, w5, w6$ of the author $a2$ are described by $11$ keywords (some can be equal)
\[ w1\colon k1, k2, k3; w3\colon k2, k3, k4, k6; w4\colon k3, k5; w6\colon k3, k6 \]
and that the keyword $k3$ is used to describe the corresponding works $w2, w3, w4, w6$ co-authored by 17 authors (some can be equal)
\[ w2\colon a1, a4, a5; w3\colon a1, a2, a3, a5, a6; w4\colon a2, a4, a5, a7, a8; w6\colon a2, a5, a7, a9 \]

A primary application of the truncated network scheme is for determining the set of the most important authors and keywords
\[ A_1 = \{ a \in A : \wod_{\m{WA}}(a) \geq t_A \}  \quad \mbox{and} \quad K_1 = \{ k \in K : \wod_{\m{WK}}(k) \geq t_K \} \]
where $t_A$ and $t_K$ are selected threshold values. Note that in the computation of $A_1$ and $K_1$ we used only the basic networks and their weighted degrees.

Ordering both vectors in decreasing order we get permutations $p_A$ and $p_K$
\[ p_A = ( 2 \ 5 \ 1 \ 3 \ 4 \ 6 \ 7 \ 8 \ 9 ) \quad \mbox{and} \quad p_K = ( 3 \ 6 \ 2 \ 5 \ 1 \ 4 ) \]
We select the largest three elements in each set
\[ A_1 = \{ a2, a5, a1 \}  \quad \mbox{and} \quad K_1 = \{ k3, k6, k2 \} \]
and compute the corresponding truncated network
\[ \mathbf{\s{AK}}_{11} = \kbordermatrix{
 & k3 & k6 & k2\\
 a2  & 3 & 2 & 2 \\
 a5  & 4 & 2 & 1 \\
 a1  & 2 & 1 & 2 \\
}\]
the corresponding vectors $\mathbf{oInt}$, $\mathbf{iInt}$, $\mathbf{oExt}$ and $\mathbf{iExt}$
\[ \mathbf{oInt} =
\kbordermatrix{
 & a2 & a5 & a1\\
 & 7 & 7 & 5\\
}  \quad \qquad \quad 
\mathbf{iInt} =
\kbordermatrix{
 & k3 & k6 & k2\\
 & 9 & 5 & 5\\
}  \]
\[  \mathbf{oExt} =
\kbordermatrix{
 & a2 & a5 & a1\\
 & 4 & 3 & 4\\
} \quad \qquad \quad
\mathbf{iExt} =
\kbordermatrix{
 & k3 & k6 & k2\\
 & 8 & 4 & 3\\
} \]
and contributions of parts of the network $\mathbf{\s{AK}}$
\[ T_{11} = \sum_{a \in A_1} \mbox{oInt}(a) = 19, \quad
   T_{10} = \sum_{a \in A_1} \mbox{oExt}(a) = 11, \quad
   T_{01} = \sum_{k \in K_1} \mbox{iExt}(k) = 15 \]
and finally, since  $T(\mathbf{\s{AK}}) = \sum_{a \in A} \wod_{AK}(a) = 53$
\[ T_{00} =  T(\mathbf{\s{AK}}) - T_{11} -  T_{10} -  T_{01} = 8 \]
Note that we used only information from $\mathbf{\m{WA}}$, $\mathbf{\s{WK}}$ and $\mathbf{\s{AK}}_{11}$.
 
The reader can check the obtained results using the reordered complete matrix $\mathbf{\s{AK}}$
\[ \mathbf{\s{AK}} = \kbordermatrix{
 & \mathbf{k3} & \mathbf{k6} & \mathbf{k2} & k5 & k1 & k4\\
 \mathbf{a2}  & \mathbf{3} & \mathbf{2} & \mathbf{2} & 2 & 1 & 1 \\
 \mathbf{a5}  & \mathbf{4} & \mathbf{2} & \mathbf{1} & 1 & 1 & 1 \\
 \mathbf{a1}  & \mathbf{2} & \mathbf{1} & \mathbf{2} & 1 & 2 & 1 \\
 a3  & 1 & 1 & 2 & 1 & 1 & 1 \\
 a4  & 2 & 0 & 0 & 1 & 1 & 0 \\
 a6  & 1 & 1 & 1 & 0 & 0 & 1 \\
 a7  & 2 & 1 & 0 & 1 & 0 & 0 \\
 a8  & 1 & 0 & 0 & 1 & 0 & 0 \\
 a9  & 1 & 1 & 0 & 0 & 0 & 0 \\
}  \]

In the following, we will analyze some special types of derived networks (projections and normalizations).

\subsection{Co-appearance network}
The \keyw{co-appearance} network $\mathbf{Co} = [co[a,b]]_{A \times A}$ can be obtained from the network $\mathbf{\m{WA}}$ as a derived network (column projection)
 \[ \mathbf{Co} = \mathbf{\m{WA}}^T \cdot \mathbf{\m{WA}}\] 
 In the case where $\mathbf{\m{WA}}$ is an authorship network the network $\mathbf{Co}$  is a co-authorship (called also collaboration) network. As we know \cite{onbib}, for $a\ne b$, $co[a,b] =$ number of works co-authored by authors $a$ and $b$. In a special case, $a=b$, we have $co[a,a] =$ number of works from $W$ written by the author $a$.

The network  $\mathbf{Co}$ is symmetric, $co[a,b] = co[b,a]$.
Therefore $\mathbf{wod}_{Co} = \mathbf{wid}_{Co}$ and by (\mbox{Oeq:} $AK \to Co$, $\mathbf{\m{WA}} \to \mathbf{\m{WA}}$, $\mathbf{\s{WK}} \to \mathbf{\m{WA}}$) we get
\[ \mathbf{wod}_{Co} = \mathbf{\m{WA}}^T \cdot \mathbf{wod}_{\m{WA}} \]


\subsubsection{Example}
For our toy network we get
\[ \mathbf{wod}_{\m{WA}} = \kbordermatrix{
 & w1 & w2 & w3 & w4 & w5 & w6 & w7\\
 & 3 & 3 & 5 & 5 & 6 & 4 & 0\\
} \]
\[ \mathbf{wod}_{Co} =  \kbordermatrix{
 & a1 & a2 & a3 & a4 & a5 & a6 & a7 & a8 & a9\\
 & 17 & 23 & 14 & 8 & 23 & 11 & 9 & 11 & 4\\
} \]
The most collaborative authors are $a2$ and $a5$ with 23 collaborations each. The author $a4$ has 8 collaborations
\[ w2\colon a1, a4, a5; w4\colon a2, a4, a5, a7, a8 \]
She/he collaborated twice with the author $a5$, but also with her/him-self.

\subsection{Normalized co-appearance network}  

The works with a large number of co-authors are "overrepresented" in the network $\mathbf{Co}$. For example, the co-authorship of authors of a paper with 2 authors counts the same as the co-authorship between any pair of authors of a paper with 1000 co-authors; a paper with 1000 co-authors adds 1000000 links to the projection network; while a single-author paper only a loop. For this reason, the number $co[a,b]$ is not the best measure for measuring the collaboration intensity.

Ideally, we would assign (a fractional approach) to each authorship $(w,a) \in L$ the weight $wa[w,a]$  expressing the proportion of the contribution of author $a$  to the work $w$ such that $\sum_a wa[w,a] = 1$ (if the work $w$ has at least one author). Unfortunately, usually, we haven't precise information about each author's contribution -- we treat all authors equally. This is a basis of the \keyw{standard normalization} $n(\mathbf{\m{WA}}) = [wan[w,a]]$ where
\[ wan[w,a] = \frac{wa[w,a]}{\max(1,\wod_{\m{WA}}(w))} \]
Note that if $\od_{\m{WA}}(w) = 0$ then $wan[w,a] = 0$, and  if $\od_{\m{WA}}(w) = 1$ then  $wan[w,a] = 1$.
We have
\[ \wod_{n(\m{WA})}(w) = \sum_{a \in A} wan[w,a] = \mbox{sign}(\od_{\m{WA}}(w))  \in \{0, 1\}\]
In binary networks it holds $\wod_{\m{WA}}(w) = \od_{\m{WA}}(w)$.

The meaning of the weighted in-degree of node $a$ in the normalized authorship network $n(\mathbf{\m{WA}})$, $\mbox{wid}_{n(\m{WA})}(a) = \sum_{w \in W} wan[w,a]$, is the \keyw{fractional contribution of the author} $a$ to all works. 

For an author $a \in A$, we can express her/his \keyw{collaborativness} \cite[p. 854]{onbib} as
\[ K(a) = 1 - \frac{\mbox{wid}_{n(\m{WA})}(a)}{\max(1,\id_{\m{WA}}(a) )} \]

The standard normalization  $n(\mathbf{\m{WA}})$ can be applied to any network with a positive weight $w: L \to \RR^+$ provided that it has a meaningful interpretation concerning our research question. For example, for an authorship network   $\mathbf{\m{WA}}$, what is the meaning of $n(\mathbf{\m{WA}}^T)$ to collaboration?

The (standard) \keyw{normalized co-appearance} network matrix $\mathbf{Cn} = [cn[a,b]] $ is obtained as a column projection of the normalized network $n(\mathbf{\m{WA}})$
\[ \mathbf{Cn} =  n(\mathbf{\m{WA}})^T \cdot n(\mathbf{\m{WA}}) \]
The normalized co-appearance network $\mathbf{Cn}$ is symmetric.

From
\[ \wid_{Cn}(a) = \wod_{Cn}(a) = \sum_{b \in A} cn[a,b] = \sum_{w \in W} wan[w,a] \cdot \sum_{b \in A} wan[w,b] \]
\[ = \sum_{w \in W} wan[w,a] \cdot  \mbox{sign}(\od_{\m{WA}}(w))  = \sum_{w \in W} wan[w,a]  =  \wid_{n(\m{WA})}(a) \]
we see that the authors have the same weighted in-degree in networks $n(\mathbf{\m{WA}})$ and $\mathbf{Cn}$
\[ \mathbf{wid}_{Cn} = \mathbf{wod}_{Cn} = \mathbf{wid}_{n(\m{WA})}\]
and therefore also equal totals
\[ T(\mathbf{Cn})   =   \sum_{a \in A} \wod_{Cn}(a)  = \sum_{a \in A} \wid_{n(\m{WA})}(a) = 
  T(n(\mathbf{\m{WA}}))\]
Let's compute
\[ T(n(\mathbf{\m{WA}})) = \sum_{w \in W} \wod_{n(\m{WA})}(w) = \sum_{w \in W} \mbox{sign}(\od_{\m{WA}}(w)) =   \sum_{w \in W_{\m{WA}}^{[1]}} 1  = | W_{\m{WA}}^{[1]}| \]
where $W_{\m{WA}}^{[d]} = \{ w \in W : \od_{\m{WA}}(w) \geq d \}$.

In the bibliometric interpretation, this means that in the normalized network $\Net_{n(\m{WA})}$, each work with at least one author has a value 1 that it is distributed over the links of the normalized co-authorship network $\Net_{Cn}$.

\subsubsection{Example}

For our toy example we get
\[ \mathbf{wid}_{n(\m{WA})} = 
\kbordermatrix{
 & a1 & a2 & a3 & a4 & a5 & a6 & a7 & a8 & a9\\
 & 1.0333 & 1.15 & 0.7 & 0.5333 & 1.15 & 0.3667 & 0.45 & 0.3667 & 0.25\\
} \]
\[ T(\mathbf{Cn})   =  T(n(\mathbf{\m{WA}})) =   \sum_{a \in A} \wid_{n(\m{WA})}(a) = | W_{\m{WA}}^{[1]}| = 6 \]
The most collaborative are authors $a2$ and $a5$ that contributed to all their co-authored works 1.15 work each.

\subsubsection{Example}

In Table~\ref{table} the effects of different thresholds $t$ on the corresponding truncated normalized co-authorship network $\mathbf{Cn}_{11}(t)$ is presented for two real-life authorship networks. It was computed using the program Pajek (see Appendix A). The \emph{iMetrics} network was analyzed in the paper \cite{iMetrics} and is an example of a typical authorship network. The network \emph{HKUST1} was created from Scopus by Nataliya Matveeva for her study of young universities \cite{HKUST1}. It contains information about papers published in the years 2017-2019 by members of HKUST (The Hong Kong University of Science and Technology).
Its number of authors per paper distribution is a combination of a ``regular'' part (typical for authorship networks) and a large number of papers with more than 1000 co-authors -- an ``irregular'' part.
The paper on the ATLAS and CMS experiments \cite{JHEP} has 5215 co-authors. The second ``largest'' paper has 5096 co-authors, and there are 295 papers with the number of co-authors in the interval 2824--2953.
This is nothing special these days \cite{hypera} -- the Guinness World Record 653537 states:
The most authors on a single peer-reviewed academic paper is 15025 and was achieved by 
the COVIDSurg and GlobalSurg Collaboratives at the University of Birmingham and 
the University of Edinburgh in the UK, as verified on 24 March 2021
\cite{Guinness}. The problem is that such a paper contributes $15025^2 = 225750625$ links with a weight $4.429667 \, 10^{-9}$ to the network  $\mathbf{Cn}$.

In Table~\ref{table}, $n$ is the number of nodes, $m$ is the number arcs, and $avdeg$ is the average degree in $\mathbf{Cn}_{11}(t)$.

Note the fast decrease of the number of links $m$ in the \emph{HKUST1} network -- the contributions of most of the authors of hyperauthored papers \cite{hypera} are very small.
\begin{table}
\caption{Truncations of iMetrics network and HKUST1 network.\label{table}}
\[
\begin{array}{l|rrr|rrr|}
        &  \multicolumn{3}{c|}{\emph{iMetrics}} &  \multicolumn{3}{c|}{\emph{HKUST1}} \\
t &      n  &   m  &  avdeg   &       n &    m &   avdeg \\ \hline
\ge 0    &  33919  &  225931  &  13.32   &   28108 & 45365272 & 3227.92 \\
\ge 1/10 &  32418  &  191888  &  11.84   &   17656 &  3216796 &  364.39 \\
\ge 1/5  &  26247  &  134049  &  10.21   &   10213 &    86529 &   16.94 \\
\ge 1/3  &  14381  &   71967  &  10.01   &    5171 &    45845 &   17.73 \\
\ge 1/2  &  12781  &   60587  &   9.48   &    4032 &    32806 &   16.27 \\
\ge 1    &   6211  &   32395  &  10.43   &    1799 &    13723 &   15.26 \\
\ge 2    &   1832  &   14900  &  16.27   &     689 &     4195 &   12.18 \\
\ge 3    &    964  &    9306  &  19.31   &     369 &     1743 &    9.45 \\
\ge 5    &    446  &    4646  &  20.83   &     172 &      646 &    7.51 \\
\ge 10   &    162  &    1450  &  17.90   &      55 &      125 &    4.55 \\ \hline
\end{array}
\]
\end{table}

\subsection{Strictly normalized co-authorship network}  

Another normalization of the authorship network was proposed by Newman \cite{strict}. It is based on the notion of strict collaboration -- self-collaboration is excluded. An author is collaborating only with different others. This has the consequence that single-author works are not considered in the analysis.

The \keyw{strict} or \keyw{Newman}'s normalization $n'(\mathbf{\m{WA}}) = [nwa'[w,a]]$ is defined by
\[ wan'[w,a] = \frac{wa[w,a]}{\max(1,\od_{\m{WA}}(w)-1)} \]
We have
\[  \wod_{n'(\m{WA})}(w) = \sum_{a \in A} wan'[w,a] = \frac{\od_{\m{WA}}(w)}{\max(1,\od_{\m{WA}}(w)-1)} \]
Note that if $\od_{\m{WA}}(w) = 0$ then $wan'[w,a] = 0$, and  if $\od_{\m{WA}}(w) = 1$ then  $wan'[w,a] = 1$.

The \keyw{strict} co-authorship network $\mathbf{Ct} = [ct[a,b]] $ is obtained as
\[ \mathbf{Ct} =  D_0(n(\mathbf{\m{WA}})^T \cdot n'(\mathbf{\m{WA}})) \]
where $D_0(\mathbf{M})$ sets the diagonal of matrix $\mathbf{M}$ to $0$. It is symmetric
\[ ct[a,b] = \sum_{w \in W} wan^T[a,w] \cdot wan'[w,b] =  \sum_{w \in W} \frac{wa[w,a]}{\max(1,\od_{\m{WA}}(w))} \cdot \frac{wa[w,b]}{\max(1,\od_{\m{WA}}(w)-1)}  = \]
 \[ =  \sum_{w \in W} \frac{wa[w,b]}{\max(1,\od_{\m{WA}}(w))} \cdot \frac{wa[w,a]}{\max(1,\od_{\m{WA}}(w)-1)}  =ct[b,a] .\]

What about the weighted out-degree of $\mathbf{Ct}$? Because by definition, $ct[a,a] = 0$, we have
\[ \wod_{Ct}(a) = \sum_{b \in A \setminus \{a\}} ct[a,b] = \sum_{w \in W} \frac{wan[w,a]}{\max(1,\od_{\m{WA}}(w)-1)} \sum_{b \in A \setminus \{a\}} wa[w,b] = \]
The matrix $\mathbf{\m{WA}}$ is binary, $wa[w,a] \in \{0.1\}$. If $wa[w,a] = 0$ the term in the $\sum_{w \in W}$ has value $0$. So we can assume $wa[w,a] = 1$. This means that $a \in oN(w)$, and $wa[w,b] = 1$ means that also $b \in oN(w)$. Therefore
\[ \sum_{b \in A \setminus \{a\}} wa[w,b] = |oN(w) \setminus \{a\}| = \od_{\m{WA}}(w) - 1 \]
Now, we can continue 
 \[ = \sum_{w \in W_{\m{WA}}^{[2]}} \frac{wan[w,a] \cdot (\od_{\m{WA}}(w)-1)}{\max(1,\od_{\m{WA}}(w)-1)}  = \sum_{w \in W_{\m{WA}}^{[2]}} wan[w,a] = \]
 \[ =  \wid_{n(\m{WA})}(a) - \sum_{w \in S} wan[w,a] =  \wid_{n(\m{WA})}(a) - | S_a | \]
where $S = \{w \in W: \od_{\m{WA}}(w) = 1\}$ is the set of single-author works and $S_a = \{ w \in S : wa[w,a] = 1 \}$ is the set of single-author works written by the author $a$. We have
\[ \wod_{Ct}(a) = \wid_{Ct}(a) =  \wid_{n(\m{WA})}(a) - | S_a | \]
For the total of $\mathbf{Ct}$ we get
\[ T(\mathbf{Ct}) = \sum_{a \in A}  \wod_{Ct}(a) = \sum_{a \in A} ( \wid_{n(\m{WA})}(a) - | S_a |) =  T(n(\mathbf{\m{WA}})) - |S| = |W_{\m{WA}}^{[1]}| - |S| = |W_{\m{WA}}^{[2]}| \]
Each work with at least two authors has a value of 1 that is in the strict co-authorship network $\mathbf{Ct}$ distributed over its links.

Again we can compute fast the weighted degrees of $\mathbf{Ct}$. Therefore the scheme used for the truncated standard fractional network can be applied also for \keyw{truncated strict fractional network}
\[ \mathbf{Ct}_{11} = \mathbf{Ct}[A_1,A_1]  = D_0(n(\mathbf{\m{WA}})[W,A_1]^T \cdot n'(\mathbf{\m{WA}})[W,A_1]) \]

\subsubsection{Example}

Because in strict co-authorship networks single-author works play a special role, we extended our toy network WA with two additional works -- the work $w8$ authored by $a6$ and the work $w9$ authored by $a2$. We labeled the new authorship network $\mathbf{\m{WA}e}$. From the weighted in-degree of its normalized network
\[ \mathbf{wid}_{n(\m{WA}e)} = \kbordermatrix{
 & a1 & a2 & a3 & a4 & a5 & a6 & a7 & a8 & a9\\
 & 1.0333 & 2.15 & 0.7 & 0.5333 & 1.15 & 1.3667 & 0.45 & 0.3667 & 0.25\\
} \]
we have to subtract $1$ from the author of each single-author work ($a6$ and $a2$) to get
\[ \mathbf{wod}_{Ct} = \kbordermatrix{
 & a1 & a2 & a3 & a4 & a5 & a6 & a7 & a8 & a9\\
 & 1.0333 & 1.15 & 0.7 & 0.5333 & 1.15 & 0.3667 & 0.45 & 0.3667 & 0.25\\
} \]
and
\[ T(\mathbf{Ct}) = \sum_{a \in A} \wod_{Ct}(a) = |W_{\m{WA}}^{[2]}| = 6 \]
Again, the most collaborative authors are $a2$, $a5$, and $a1$.
 
\subsection{Product of normalized networks}
For the \keyw{product of normalized networks}
\[ \mathbf{AKn} = n(\mathbf{\m{WA}})^T \cdot n(\mathbf{\s{WK}}) \]
we get from (Oeq, Ieq: $AK \to AKn$, $\mathbf{\m{WA}} \to n(\mathbf{\m{WA}})$, $\mathbf{\s{WK}} \to n(\mathbf{\s{WK}})$)
\[ \mathbf{wod}_{AKn} = n(\mathbf{\m{WA}})^T \cdot \mathbf{wod}_{n(\s{WK})} \quad \mbox{and}
\quad \mathbf{wid}_{AKn} = n(\mathbf{\s{WK}})^T \cdot \mathbf{wod}_{n(\m{WA})} \]
and for its total
\[ T(\mathbf{AKn}) = \mathbf{wod}_{\m{WA}n} \cdot \mathbf{wod}_{\s{WK}n}  = \sum_{w \in W} \mbox{sign}(\od_{\m{WA}}(w)) \cdot  \mbox{sign}(\od_{\m{WK}}(w)) = \]
\[ = | \{ w \in W : (\od_{\m{WA}}(w) > 0) \land (\od_{\m{WK}}(w) > 0) | = | W_{\m{WA}}^{[1]} \cap W_{\s{WK}}^{[1]}  | \]
In bibliometrics terms, each work with at least one author and at least one keyword has a value of 1 which is in the network $\mathbf{AKn}$ distributed over its links.

\subsubsection{Example}

For our toy networks we have
\[ \mathbf{wod}_{n(\m{WA})} = \kbordermatrix{
 & w1 & w2 & w3 & w4 & w5 & w6 & w7\\
 & 1 & 1 & 1 & 1 & 1 & 1 & 0\\
} \]
\[ \mathbf{wod}_{n(\s{WK})} = \kbordermatrix{
 & w1 & w2 & w3 & w4 & w5 & w6 & w7\\
 & 1 & 1 & 1 & 1 & 0 & 1 & 1\\
} \]
\[ \mathbf{wod}_{AKn} = \kbordermatrix{
 & a1 & a2 & a3 & a4 & a5 & a6 & a7 & a8 & a9\\
 & 0.8667 & 0.9833 & 0.5333 & 0.5333 & 0.9833 & 0.2 & 0.45 & 0.2 & 0.25\\
} \]
\[ \mathbf{wid}_{AKn} = \kbordermatrix{
 & k1 & k2 & k3 & k4 & k5 & k6\\
 & 0.8333 & 0.5833 & 1.75 & 0.25 & 0.8333 & 0.75\\
} \]

\[ T(\mathbf{AKn}) =  5 \]

\subsection{Linking through a network}

Assume that another network $\mathbf{S}$ on $W \times W$ is given. The network
\[ \mathbf{Q} =  n(\mathbf{\m{WA}})^T \cdot \mathbf{S} \cdot n(\mathbf{\m{WA}}) \]
links nodes from the set $A$  \keyw{through} the network $\mathbf{S}$. From \cite{fraction} we know that
\begin{itemize}
\item If $\mathbf{S}$ is symmetric, $\mathbf{S}^T = \mathbf{S}$, then also $\mathbf{Q}$ is symmetric, $\mathbf{Q}^T = \mathbf{Q}$.
\item if $W_{\m{WA}}^{[1]} = W$ then $T(\mathbf{Q}) = T(\mathbf{S})$.
\end{itemize}

Let's look at 
\[ \wod_{Q}(a) = \sum_{b \in A} q[a,b] = \sum_{b \in A} \sum_{w \in W} \sum_{z \in W} wan[w,a] \cdot s[w,z] \cdot wan[z,b] = \]
\[ = \sum_{w \in W} \sum_{z \in W} wan[w,a] \cdot s[w,z] \cdot \mbox{sign}(\od_{\m{WA}}(z))  = 
\sum_{w \in W}  wan[w,a] \cdot \sum_{z \in W_{\m{WA}}^{[1]}} s[w,z]   = \]
\[  =  \sum_{w \in W}  wan[w,a] \cdot \mbox{wod}_{S}^{[1]}(w)   \]
where $\mbox{wod}_{S}^{[1]}(w) = \sum_{z \in W_{\m{WA}}^{[1]}} s[w,z]$, or in a vector form
\[ \mathbf{wod}_{Q} = n(\mathbf{\m{WA}})^T  \cdot \mathbf{wod}_{S}^{[1]}\]

The most active authors are
\[ A_1 = \{ a \in A : \wod_{Q}(a) \geq t \} \]
Again the truncation scheme can be applied.

\subsubsection{Example}

Taking for network $\mathbf{S}$ the normalized co-citation network $\mathbf{coCin} = n(\mathbf{Ci})^T \cdot n(\mathbf{Ci})$ (see Subsection 4.7) we get the normalized co-citation network between authors
\[ \mathbf{coCan} =  n(\mathbf{\m{WA}})^T \cdot \mathbf{coCin} \cdot n(\mathbf{\m{WA}}) \]
Computing its weighted out-degree we get
\[ \mathbf{wod}_{coCan} = \kbordermatrix{
 & a1 & a2 & a3 & a4 & a5 & a6 & a7 & a8 & a9\\
 & 0.4093 & 0.8676 & 0.2981 & 0.2222 & 0.9787 & 0.2981 & 0.5694 & 0.3426 & 0.4583\\
} \]
It can be computed more efficiently by first determining $\mathbf{wod}_{S}^{[1]} = \wod(\mathbf{coCin}_{W\times W_{\m{WA}}^{[1]}} )$
\[ \mathbf{wod}_{S}^{[1]} = \kbordermatrix{
 & w1 & w2 & w3 & w4 & w5 & w6 & w7\\
 & 0 & 0.3333 & 0.3333 & 0.5556 & 1.3889 & 1.8333 & 0.2222\\
} \]
and afterward applying the equality $\mathbf{wod}_{Q} = n(\mathbf{\m{WA}})^T  \cdot \mathbf{wod}_{S}^{[1]}$ the same result
\[ \mathbf{wod}_{Q} = \kbordermatrix{
 & a1 & a2 & a3 & a4 & a5 & a6 & a7 & a8 & a9\\
 & 0.4093 & 0.8676 & 0.2981 & 0.2222 & 0.9787 & 0.2981 & 0.5694 & 0.3426 & 0.4583\\
} \]

\subsection{Co-citation, authors co-citation and bibliographic coupling}

The \keyw{co-citation} network is defined as the column projection of the citation network
\[ \mathbf{coCi} = \mathbf{Ci}^T\cdot\mathbf{Ci} \]
and the \keyw{normalized co-citation} network as  the column projection of the normalized citation network ($\mathbf{Cin} = n(\mathbf{Ci})$) 
\[ \mathbf{coCin} = \mathbf{Cin}^T\cdot\mathbf{Cin}. \]
Both $\mathbf{coCi}$ and $\mathbf{coCin}$ are symmetric. They follow patterns discussed in Subsections 4.2 and 4.3.

\keyw{Normalized authors co-citation} network is obtained by linking authors through the normalized co-citation network
\[ \mathbf{coCan} =  n(\mathbf{\m{WA}})^T \cdot \mathbf{coCin} \cdot n(\mathbf{\m{WA}}) = \]
\[ = n(\mathbf{\m{WA}})^T \cdot \mathbf{Cin}^T\cdot\mathbf{Cin} \cdot n(\mathbf{\m{WA}}) = 
 \mathbf{Can}^T\cdot\mathbf{Can} \]
where $\mathbf{Can} = \mathbf{Cin}\cdot n(\mathbf{\m{WA}}) $. \medskip

Applying the result from Subsection 4.6, we get
\[  \mathbf{wod}_{coCan} = n(\mathbf{\m{WA}})^T  \cdot \mathbf{wod}_{coCin}^{[1]} \]
that can be expanded into 
\[ \wod_{coCin}^{[1]}(w) = \sum_{z \in W_{\m{WA}}^{[1]}} coCin[w,z] = \]
\[ = \sum_{t \in W} Cin[t,w] \sum_{z \in W_{\m{WA}}^{[1]}} Cin[t,z] = \sum_{t \in W} Cin[t,w] \wod_{Cin}^{[1]}(t) \]
or in a vector form
\[  \mathbf{wod}_{coCin}^{[1]} = \mathbf{Cin}^T  \cdot \mathbf{wod}_{Cin}^{[1]} \]
Therefore the vector $\mathbf{wod}_{coCan}$ can be faster computed using the relation
\[ \mathbf{wod}_{coCan} = n(\mathbf{\m{WA}})^T \cdot (\mathbf{Cin}^T \cdot \mathbf{wod}_{Cin}^{[1]}) \]
It is easy to see that also
\[  \mathbf{wid}_{Can} = n(\mathbf{\m{WA}})^T  \cdot \mathbf{wid}_{n(Ci)} = \mathbf{wod}_{coCan}\]

The \keyw{bibliographic coupling} network is defined as the row projection of the citation network
\[ \mathbf{biCo} = \mathbf{Ci}\cdot\mathbf{Ci}^T  .\]  
It is symmetric. From Oeq ($AK \to biCo$, $\mathbf{\m{WA}} \to \mathbf{Ci}^T$, $\mathbf{\s{WK}} \to \mathbf{Ci}^T$) we get
\[ \mathbf{wod}_{biCo} = \mathbf{Ci}  \cdot \mathbf{wid}_{Ci}\]

Because the normalization $n(\mathbf{Ci}^T)$ makes no sense, the fractional approach can not be directly applied to bibliographic coupling. The selection of important nodes for the solutions proposed in  \cite{fraction} is still to be elaborated.

\subsubsection{Example}

For our toy networks, we get
\[ \mathbf{wod}_{Cin}^{[1]} = \kbordermatrix{
 & w1 & w2 & w3 & w4 & w5 & w6 & w7\\
 & 1 & 1 & 0.6667 & 1 & 1 & 0 & 0\\
} \]
\[ \mathbf{wod}_{coCin}^{[1]} = \kbordermatrix{
 & w1 & w2 & w3 & w4 & w5 & w6 & w7\\
 & 0 & 0.3333 & 0.3333 & 0.5556 & 1.3889 & 1.8333 & 0.2222\\
} \]
\[ \mathbf{wod}_{coCan} = \kbordermatrix{
 & a1 & a2 & a3 & a4 & a5 & a6 & a7 & a8 & a9\\
 & 0.4093 & 0.8676 & 0.2981 & 0.2222 & 0.9787 & 0.2981 & 0.5694 & 0.3426 & 0.4583\\
} \]




\section{Conclusions}

Although the notion of truncated networks was presented in the context of bibliographic networks the results can be applied also in other fields described by collections of networks.

For experimenting with smaller networks (up to 1000 nodes in each set) we developed a collection of R functions \texttt{\textbf{bibmat}} that supports introduced operations on networks represented by matrices. For analyzing large networks we can use their implementation in Pajek as Pajek's commands or macros.


\section*{Acknowledgments}

\ifblind 
\textbf{Acknowledgments omitted in the blind version.}
\else
The computational work reported in this paper was performed using a collection of R functions \texttt{\textbf{bibmat}} and the program Pajek for analysis of large networks \cite{pajek}. The code and data are available at Github/Bavla \cite{bibR}. 

This work is supported in part by the Slovenian Research Agency
 (research program P1-0294, research program CogniCom (0013103)
at the University of Primorska, and research projects J5-2557, J1-2481, and J5-4596),
 and prepared within the framework of the COST action CA21163 (HiTEc).

\fi


\clearpage
\appendix
\section{Computing truncated normalized co-authorship network at level $t$ in Pajek}
\begin{lstlisting}
read Network [WA_HKUST1.net]
Networks Info Button -> Rows = 9225; Cols = 28108
Macro/Play [norm2.mcr][9225] -> normalized network WAn
Network/Create vector/Centrality/Weighted degree/Input -> wid_WAn
Network/2-Mode network/Partition into 2 modes
Operations/Vector+Partition/Extract subvector [2]
Vectors Info Button [0][0.09999 0.49999 0.99999 1.99999 2.99999 4.99999 9.99999]
select a threshold t (= 0.99999)
Vector/Create scalar/Sum
File/Vector/Change label [T(WAn)]
select wid_WAn as the First vector
Vector/Make partition/by intervals/Selected thresholds [t] -> one
Operations/Vector+Partition/Extract subvector [2]
File/Vector/Change label [wid_WAn-one]
Partition/Create constant partition [9225,0]
select partition one as the Second
Partitions/Fuse partitions
Operations/Network+Partition/Extract/Subnetwork induced [0,2]
Network/2-Mode network/Partition into 2 modes
Partition/Binarize partition [2]
Network/Create partition/degree/output
select the binarized partition as the Second
Partitions/Add (First+Second)
Operations/Network+Partition/Extract/Subnetwork induced [1-*]
File/Network/Change label [WAn1]
Network/2-Mode network/2-mode to 1-mode/include loops [On]
Network/2-Mode network/2-mode to 1-mode/columns
Network/Create vector/Centrality/weighted degree/output -> alpha
Vector/Create scalar/Sum
File/Vector/Change label [T11]
select wid_WAn as the First vector
select alpha as the Second vector
Vectors/Subtract(First-Second)
Vector/Create scalar/Sum
File/Vector/Change label [T10]
select T(WAn) as the First vector
select T11 as the Second vector
Vectors/Subtract(First-Second)
select T10 as the Second vector
Vectors/Subtract(First-Second)
Vectors/Subtract(First-Second)
File/Vector/Change label [T00]
remove auxiliary vectors, partitions, networks
\end{lstlisting}

\end{document}